\documentclass[useAMS,usenatbib]{mn2e} 
\usepackage{epsfig,float}
\usepackage{graphicx,amsmath,multicol}
\usepackage{psfrag}
\usepackage[dvipsnames]{xcolor}
\definecolor{webgreen}{rgb}{0,.5,0}
\definecolor{webbrown}{rgb}{.6,0,0}
\usepackage[pdfpagelabels]{hyperref}
\hypersetup{%
   colorlinks=true,hyperfootnotes=false,%
   breaklinks=true,%
   plainpages=false, bookmarksnumbered, bookmarksopen=true,
   bookmarksopenlevel=1,%
   urlcolor=webbrown, linkcolor=webbrown, citecolor=webgreen,
   }
   
\newcommand{\comments}[1]{}
\newcommand       \be           {\begin{equation}}
\newcommand       \ee           {\end{equation}}
\newcommand       \ba           {\begin{eqnarray}}
\newcommand       \ea           {\end{eqnarray}}

\newcommand       \apj          {ApJ}
\newcommand       \apjl         {ApJL}

\newcommand       \mnras        {MNRAS}
\newcommand       \araa         {Ann. Rev. Astr. Astr.}
\def\sun{\odot}
\def\msun{\rm \ M_\odot}

\def\lesssim{\mathrel{\hbox{\rlap{\hbox{\lower4pt\hbox{$\sim$}}}\hbox{$<$}}}}
\def\gtrsim{\mathrel{\hbox{\rlap{\hbox{\lower4pt\hbox{$\sim$}}}\hbox{$>$}}}}

\defcitealias{sha12a}{Paper~I}

\title[Thermal Conduction \&  Multiphase Gas in Cluster Cores]
{Thermal Conduction and Multiphase Gas in Cluster Cores}  
\author[B.\ Wagh, P.\ Sharma, M. McCourt]
{Baban Wagh$^\dag$, Prateek Sharma$^\dag$, Michael McCourt$^\ddag$ \\
$^\dag$Department of Physics and Joint Astronomy Program, Indian Institute of Science, Bangalore, India 560012  (baban,prateek@physics.iisc.ernet.in) \\
$^\ddag$ Astronomy Department and Theoretical Astrophysics Center, University of California, Berkeley, CA 94704 USA (mkmcc@berkeley.edu)\\}

\begin{document}

\pagerange{\pageref{firstpage}--\pageref{lastpage}} \pubyear{2013}
\maketitle
\label{firstpage}

\begin{abstract}
We examine the role of thermal conduction and magnetic fields in cores of galaxy clusters through global simulations of the intracluster medium (ICM). In particular, we study the influence of thermal conduction, both isotropic and anisotropic, on the condensation of multiphase gas in cluster cores.
Previous hydrodynamic simulations have shown that cold gas condenses out of the hot ICM in thermal balance only when the ratio of the cooling time ($t_{\rm cool}$) and the free-fall time ($t_{\rm ff}$) is less than $\approx 10$. Since thermal conduction is significant in the ICM and it suppresses local cooling at small scales, it is imperative to include thermal conduction in such studies. We find that anisotropic (along local magnetic field lines) thermal conduction does not influence the condensation criterion for a general magnetic geometry, even if thermal conductivity is large. However, with isotropic thermal conduction cold gas condenses only if conduction is suppressed (by a factor $\lesssim 0.3$) with respect to the Spitzer value. 
\end{abstract}


\section{Introduction}

The radiative cooling time for the X-ray emitting gas in galaxy clusters is inversely proportional to the gas density, and as the inner regions of galaxy clusters have high electron densities ($\sim 0.1$ cm$^{-3}$), the cooling time is short compared to the cluster age ($\sim$Hubble time). However, the signatures of cooling and cold gas predicted due to such short cooling times are in stark contrast with observations in different wavebands (\citealt{pet03,hic05,ode08}). The discrepancy is the so called cooling flow problem (\citealt{fab94}). 

The lack of cooling in cluster cores suggests an efficient energy injection mechanism that offsets cooling losses. Some plausible heat sources are jets/bubbles powered by central AGN (Active Galactic Nuclei -- accretion powered supermassive black holes; for a review, see \citealt{mcn07}), thermal conduction from large radii (e.g., \citealt{zak03}), heating by dynamical friction and turbulence driven by infalling galaxies and subhalos (e.g., \citealt{kim05,den05}). The most promising of these sources is AGN jets and bubbles. Observations indicate that the AGN blown radio bubbles and X-ray cavities (\citealt{bir04}) -- indicative of mechanical energy input in central regions of galaxy clusters -- have power comparable to the core X-ray luminosity. Although non-feedback sources may be energetically sufficient, the core must be heated by a globally stable mechanism such as AGN feedback (stellar feedback is insufficient) that can keep up with efficient cooling.

The plasma in the ICM is magnetized with the magnetic field strength of order $1-10$ $\mu$G (\citealt{car02}). While the plasma $\beta$ (the ratio of plasma pressure to magnetic pressure) is high and the magnetic force on large scales is negligible, magnetic fields fundamentally alter the transport properties of the ICM. The Coulomb mean free path in the core is $\lesssim 0.1$ kpc,  $\sim 10$ to 12 orders of magnitude larger than the proton/electron Larmor radius. Therefore, particles move and diffuse along the local magnetic field direction. Thus, heat and momentum transport is  only along the local magnetic field lines. Given that the mean free path is much smaller than the core size ($\sim$ 10 kpc), we use Braginskii conductivity along magnetic field lines and do not consider the effects of small scale instabilities (e.g., \citealt{sch05}). The question of how small scale instabilities affect dynamics and thermodynamics at macroscopic scales is not fully resolved and is subject of intense research (e.g., \citealt{mog13}). However, the simplest interpretation of the solar wind data suggests that it is fine to use a collisional (Spitzer-H\"arm) value of thermal conductivity for scales as big as one-third of the mean free path \citep[][]{bal13}.

Self-consistent simulations with magnetic fields and anisotropic thermal conduction are especially important given the wide range of theoretical predictions for the effective conductivity of the ICM \citep[e.g.,][]{rec78,cha98,nar01}. We only consider anisotropic thermal conduction but not anisotropic viscosity; the viscous diffusion time is $\sim 10$ times longer than the thermal diffusion timescale (few 100 Myr) and the viscous effects are expected to be subdominant (\citealt{par12}). See, however, \citealt{kun12} who emphasized that Braginskii viscosity corresponds to a large pressure anisotropy that excites the Larmor radius scale mirror and firehose instabilities, and as stressed earlier, their influence on macroscopic scales is theoretically not well understood. As with thermal conduction, we choose the simplest model for the ICM viscosity; namely, we ignore it as it is small. We emphasize that while the viscosity parallel to field lines is small, the perpendicular viscosity is completely negligible. Therefore, perpendicular velocity is totally unaffected by viscosity, unlike in an isotropic fluid.

Anisotropic conduction in the core where the temperature increases with radius leads to a buoyancy instability (the heat-flux driven buoyancy instability, HBI; \citealt{qua08}) that rearranges magnetic field lines perpendicular to the radial direction, thus shutting off thermal transport to the core from outer radii (\citealt{par09}). In addition to affecting the global thermal properties, anisotropic thermal conduction also affects the local thermal instability in the core. With isotropic thermal conduction, scales smaller than the Field length (scale at which the heat diffusion timescale equals the thermal instability timescale) do not condense (\citealt{fie65}). However, if thermal conduction is anisotropic, thermal instability is not suppressed perpendicular to the field lines and long filaments directed along magnetic field lines, at scales larger than the Field length, can condense (\citealt{sha10}; more so if magnetic field lines are  tangled or form closed loops).

Motivated by the observed thermal balance in cluster cores, recent hydrodynamic simulations have studied cluster cores in rough thermal balance (\citealt{mcc12,sha12a,gas12}). These simulations, using different codes and setups, have found that local thermal instability can lead to the formation of multiphase gas only if the ratio of the cooling time ($t_{\rm cool}\approx t_{\rm TI}$ the thermal instability timescale for a constant heating rate per unit volume) and the free-fall time ($t_{\rm ff} \equiv [2r/g]^{1/2}$; $g$ is the acceleration due to gravity) is smaller than a critical value. This critical value is $ \approx 10 $ for spherical systems such as galaxy clusters. The observations of extended cold atomic filaments in cluster cores are consistent with this criterion (Fig. 11 in \citealt{mcc12}). Moreover, the $t_{\rm cool}/t_{\rm ff} \sim 10$ state of the hot gas is expected to provide a rough upper limit on the density of hot gas in halos with different masses (\citealt{sha12b}).\footnote{We note that Fig. 11 in \citet{mcc12} overestimates the $t_{\rm cool}/t_{\rm ff}$ ratio by a factor of two; given the approximations in our simulations and uncertainties in interpreting X-ray data, however, the existence of a threshold is more important than its precise value.}   

Most of the previous work did not include thermal conduction and magnetic fields. Given the importance of local thermal instability and a large thermal conductivity in the magnetized ICM, studies of cooling and heating with thermal conduction and magnetic fields  are warranted. We do just this in the present paper. Our main result is that while isotropic thermal conduction at the Spitzer value suppresses condensation  in the core even when $t_{\rm cool}/t_{\rm ff} \lesssim 10$, this criterion is essentially unchanged with the more realistic anisotropic thermal conductivity. Thus, the implications of hydrodynamic models (\citealt{sha12a, sha12b}) should hold for the magnetized plasma in massive dark matter halos.

This paper  is  organized as follows. Section 2 is a description of our numerical setup, largely based on \citealt{sha12a} (hereafter \citetalias{sha12a}) but including magnetic fields and thermal conduction. In section 3 we describe our results. In section 4 we discuss the astrophysical implications of our study. 

\section{Numerical Setup}

This paper extends the work in \citetalias{sha12a} by including magnetic fields and thermal conduction (both isotropic and anisotropic). Like that work, the background gravitational potential is given by a static NFW profile  (\citealt{nav97}). Most of our runs use a cluster-mass halo with $M_0=3.8 \times 10^{14} \msun$ and $r_s=390$ kpc (see Eq. [4] in \citetalias{sha12a}; group and massive cluster runs presented at the end correspond to $M_0=3.8 \times 10^{13}$ and $10^{15} \msun$, respectively). The initial conditions consist of the ICM in hydrostatic equilibrium but with small (maximum $\delta \rho \approx 0.3$) isobaric density perturbations. The entropy profile of the ICM (following \citealt{cav09}; $K \equiv T_{\rm keV}/n_e^{2/3}$ has a core entropy $K_0$) is specified as the initial condition. Unless specified otherwise, all parameters are identical to \citetalias{sha12a}.

We use the {\tt ZEUS-MP} code (\citealt{hay06}) in spherical geometry for our cluster simulations. The inner and outer radii for cluster simulations are 1 kpc and 200 kpc, respectively; for other halos the inner and outer radii are scaled with the halo mass. Outflow boundary conditions are used at the inner radial boundary and reflective boundary conditions are applied at the outer radial boundary. This outer radial boundary condition is different from \citetalias{sha12a} and found to be more robust than the inflow boundary condition used there.  Outer boundary conditions should not affect the results since the halo gas is in thermal balance and multiphase gas condenses much inside the outer boundary.  Reflective (periodic) boundary conditions are applied in the meridional (azimuthal) direction.

Since magnetic field geometry in galaxy clusters is largely unconstrained, we try split-monopole ($|B_r| \propto 1/r^2$), azimuthal ($B_\phi \propto 1/r$), and tangled initial magnetic field geometries. The tangled (divergenceless) magnetic field is generated by taking the curl of a vector potential. The cartesian components of the vector potential are chosen to have Fourier amplitudes $A_i(k) \propto k^{-8/3}$, where $k$ is the absolute value of the wavenumber. The minimum and maximum  wavenumbers in each direction are $2\pi/ (100~{\rm kpc})$ and $2 \pi/ (40~{\rm kpc})$, respectively. The Fourier components are added with random phases. The magnitude of the vector potential is smoothly reduced toward  poles ($\theta=0,~\pi$) and outer radii (mainly to avoid magnetically dominated regions close to poles that can severely reduce the CFL timestep; exact form of magnetic field structure should not affect the results). The field strength at 1 kpc is chosen to be $\sim 1 \mu$G, in rough agreement with the observed field strengths. The radial magnetic field runs are carried out using 2-D axisymmetric simulations with a resolution of $512 \times 256$. The tangled magnetic field simulations are carried out in 3-D with a resolution of $128\times64\times32$.\footnote{The tangled field simulations must be carried out in 3-D because tangled fields in 2-D give rise to magnetically closed flux loops which are much rarer in 3-D. Moreover, these closed field loops are thermally insulated from rest of the plasma if thermal conduction is along magnetic field lines.} Since the results of azimuthal magnetic field simulations are very similar to the simulations without thermal conduction (as expected and as shown in \citealt{mcc12}), we do not discuss them further.

Radiative cooling is included using the cooling function in Eq. (12) of \citet{sha10}. Observations and  simulations with uniformly distributed feedback heating in \citetalias{sha12a} (see also jet simulations of \citealt{gas12}) suggest that  cluster cores reach approximate thermal balance. Therefore, we impose strict thermal balance at all radii using a heating term (which mimics feedback heating and depends on radius and time; see Eq. [6] in \citetalias{sha12a}) in every radial shell. This setup  is very idealized, but the model of local thermal instability in global thermal balance captures some of the key features of cool core clusters; e.g., no cold gas can condense out of the hot ICM if the cooling time is too long compared to the free-fall time. Unlike in \citetalias{sha12a}, heating/cooling due to thermal conduction is included in imposing thermal balance. (Thus, heating term can, in principle, be negative! However, this is not very common since radiative cooling dominates over conductive heating.) This heating term corresponds to a constant heating rate per unit volume. We do not try any other form of the microscopic heating rate because this choice results in $t_{\rm cool} \approx t_{\rm TI}$ and a good match with observations (see Fig. 11 in \citealt{mcc12}).

We include thermal conduction, both isotropic and anisotropic (along local magnetic field direction). The thermal conductivity ($\propto T^{5/2}$) is normalized to the Spitzer value (Eq. 11 in \citealt{sha10}). Anisotropic thermal conduction along magnetic field lines is implemented explicitly using the method of \citet{sha07}. Thermal conduction is implemented via subcycling; we reduce the local conductivity (if needed) such that the number of subcycles  at any time is not too high compared to its initial value. A high number of subcycles means that the conductive time across the grid cell is much shorter than the sound crossing time; this ordering (and thus the dynamical significance of conduction) is maintained even when we locally limit conductivity. The conduction step becomes prohibitively slow if we do not reduce the local conductivity because very hot plasma is created intermittently when cold gas condenses out. Thermal conduction must be implemented implicitly if we want to perform high resolution simulations with realistic thermal conduction (e.g., \citealt{sha11,mey12}); we plan to implement this in future. The equations solved are the standard MHD equations with cooling, heating, and thermal conduction; namely, Eqs. 1(a,b$^\prime$,c$^\prime$,d$^\prime$) and Eq. 28 in \citet{mcc12}. 

\section{Results}

\begin{figure*}
\begin{center}
\psfrag{A}[cc][][1.2][0]{aniso. cond.}
\psfrag{I}[cc][][1.2][0]{iso. cond.}
\psfrag{N}[cc][][1.2][0]{no cond.}
\psfrag{R}[cc][][1.2][0]{$r({\rm kpc})$}
\psfrag{Y}[cc][][1.2][90]{$r({\rm kpc})$}
\psfrag{M}[cc][][1.2][0]{$\log_{10} n_e$}
\includegraphics[scale=0.5]{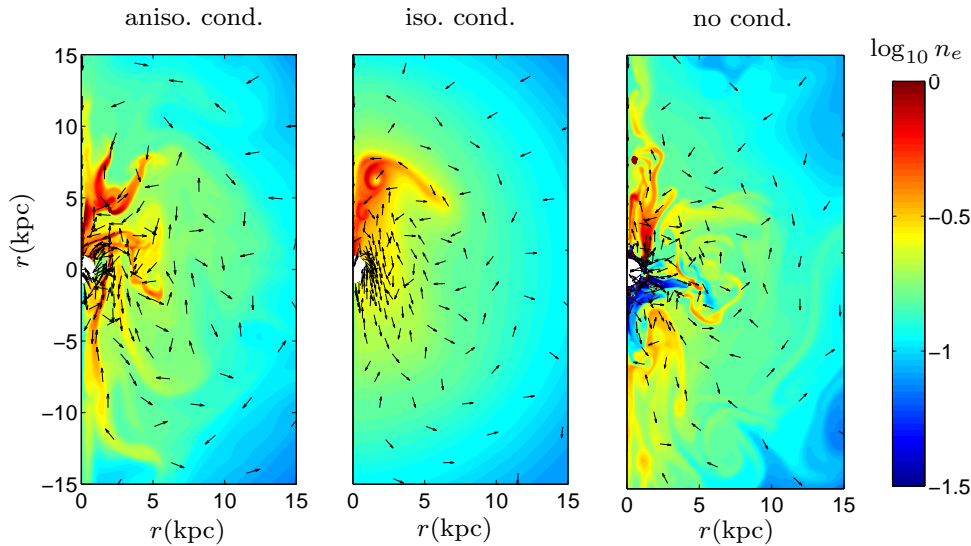}
\caption{Contour plots of electron number density just when cold gas condenses in different 2-D simulations with initial core entropy of $K_0=5$ keV cm$^2$. The number density is cut-off at the lowest and highest limits. Arrows represent magnetic field unit vectors. The cold gas is  more filamentary in presence of magnetic fields, as can be seen by comparing with Figure 4 in \citetalias{sha12a}. \label{fig:2D}}
\end{center}
\end{figure*}

Subthermal magnetic fields do not influence the linear thermal instability (\citealt{sha10}). Our key result is that isotropic thermal conduction 
suppresses multiphase gas formation but cold gas condenses with anisotropic conduction even with few times the Spitzer value, as long as  $t_{\rm cool}/t_{\rm ff} \lesssim 10$. Thus, the inferences from previous hydrodynamic simulations (e.g., \citetalias{sha12a}, \citealt{mcc12}) carry over with realistic anisotropic thermal conduction.

Figure \ref{fig:2D} shows a comparison of gas density just when cold gas condenses in simulations with anisotropic and isotropic conduction (Spitzer fraction $f=1$), and with no conduction. The initial core entropy $K_0=5$ keV cm$^2$. We  show snapshots of 2-D simulations because 3-D simulations have low resolution; however, the trend of integrated quantities (such as mass accretion rate) is similar for 2-D and 3-D runs (c.f. Fig. \ref{fig:cold_fr}). The most striking observation is that the appearance of cooler/denser gas is similar in cases with anisotropic conduction and with no conduction. The density (and temperature) is much more diffuse with isotropic conduction. The denser gas is aligned with the local magnetic field in case with anisotropic conduction. This correlation remains to some degree in simulations with isotropic conduction and without conduction because of flux freezing during buoyant motions (e.g., \citealt{kom13}). Just the visual similarity of the run with anisotropic conduction and the run without conduction (and not of  the isotropic conduction run) suggests that anisotropic conduction does not affect cold gas condensation in cool core clusters. The appearance of numerous small-scale optical and soft X-ray filaments in nearby clusters (e.g., \citealt{san07,wer10}) hints that thermal conduction in cool cluster cores is suppressed or is anisotropic.

\begin{figure}
\begin{center}
\psfrag{A}[cc][ ][1.2][90]{$\frac{t_{\rm cool}}{t_{\rm ff}}$}
\psfrag{B}{$r({\rm kpc})$}
\psfrag{C}{$r({\rm kpc})$}
\psfrag{D}{time (Gyr)}
\psfrag{E} [cc][ ][.9][90]  {$\dot{M} (M_\sun {\rm yr}^{-1})$}
\psfrag{F}{aniso. cond.} 
\psfrag{G}[cc]{     \hspace{0.15in}  iso. cond.}
\includegraphics[scale=0.35]{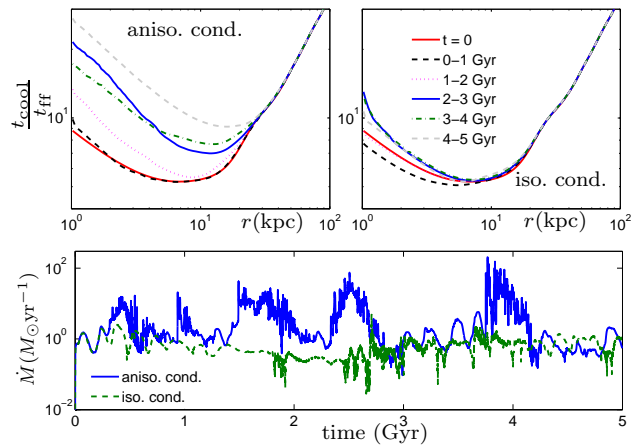}
\caption{Time- and angle-averaged radial profiles of ($t_{\rm cool}/t_{\rm ff}$) for hot ($\geq 0.1$ keV) gas in 3-D simulations with anisotropic (left) and isotropic (right) thermal conduction.
The initial core entropy is $K_0 = 5$ keV cm$^2$.  Substantial cold gas forms and drops out with anisotropic conduction, leaving behind a core with $t_{\rm cool}/t_{\rm ff} \approx 10$ at late times. Profiles are only slightly adjusted for the run with isotropic thermal conduction. Bottom panel shows the mass accretion rate through the inner radius (1 kpc) as a function of time for the two runs. Large spikes correspond to enhanced accretion in the cold phase. \label{fig:icm_reg}}
\end{center}
\end{figure}

\begin{figure}
\centering
\psfrag{B}[cc][ ][1][0]  {$K_0 ({\rm keV cm}^2)$}
\psfrag{C}{$t_{\rm cool}/t_{\rm ff}$}
\psfrag{A}[cc][ ][1][90]{$\dot{M}_{\rm cold} (M_\sun {\rm yr}^{-1})$}
\psfrag{T3kA1}{3-D aniso. cond.}
\psfrag{R2kA1}{2-D aniso. cond.}
\psfrag{R2kI1}{2-D iso. cond.}
\psfrag{T3kI1}{3-D iso. cond.}
\psfrag{R2kA10}{2-D aniso. cond.}
\psfrag{T3}{3-D no cond.}
\includegraphics[scale=0.46]{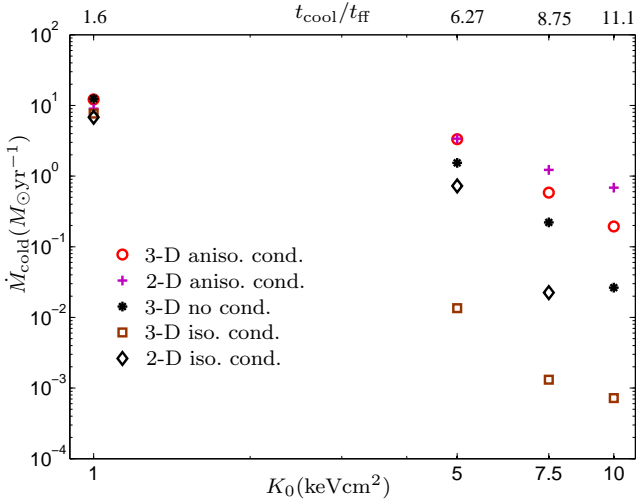}
\caption{Average mass accretion rate through the inner boundary in the cold ($< 0.01$ keV) phase as a function of the initial core entropy $K_0$ (and initial minimum $t_{\rm cool}/t_{\rm ff}$) for different cluster runs. Anisotropic conduction does not affect the cold gas formation criterion. In contrast, cold gas is suppressed with isotropic conduction for $K_0 \gtrsim 5$ keV cm$^2$. All runs with conduction use a Spitzer fraction $f=1$. \label{fig:cold_fr}}
\end{figure}

Figure \ref{fig:icm_reg} shows the time- and angle-averaged 1-D profiles of the ratio of the cooling time and the free-fall time ($t_{\rm cool}/t_{\rm ff}$) for 3-D simulations with an initial core entropy of $K_0=5$ keV cm$^2$. The left panel shows the profiles for the simulation with anisotropic conduction at the Spitzer value. The right panel shows the same  for isotropic conduction at the Spitzer value. The minimum $t_{\rm cool}/t_{\rm ff} \approx 6$ in the initial condition. Since $t_{\rm cool}/t_{\rm ff} < 10$ in the core, cold gas condenses with anisotropic conduction because of local thermal instability. Initially cold gas condenses out from the hot phase, lowering the density of the remaining hot gas. This raises the $t_{\rm cool}/t_{\rm ff} $ ratio until it reaches about 10. After that point we no longer see cold gas. Thus, we conclude that the criterion for the formation of multiphase gas and the self regulation of core entropy are essentially unaffected in presence of magnetic fields and {\em anisotropic} conduction. Compare this with the isotropic run where only a minor readjustment of the profiles happens. The bottom panel of Figure \ref{fig:icm_reg}  shows the mass accretion rate as a function of time for the same runs. Clearly, the mass accretion rate, especially in the cold phase (corresponding to spikes), is much larger with anisotropic conduction. This is consistent with the drop-out of large amount of gas with anisotropic conduction. While $K_0=5$ keV cm$^2$ does not show extensive cold gas with isotropic thermal conduction, anisotropic runs are very similar to hydrodynamic runs. Thus, we conclude that isotropic conduction at the Spitzer value is  ruled out by observations which show cold gas for clusters with core entropy $K_0 \lesssim$ 10 keV cm$^2$ (\citealt{cav08}). In contrast, anisotropic thermal conduction is consistent with these observations.

Figure \ref{fig:cold_fr} shows the mass accretion rate in the cold phase as a function of the initial core entropy ($K_0$) for different runs. The 3-D control runs without thermal conduction are included for comparison. The 2-D and 3-D runs with anisotropic and isotropic thermal conduction at the Spitzer value are also shown. Thermal conduction is expected to suppress cold phase formation, and this is evident with isotropic conduction. The runs with anisotropic conduction are not affected much and show cold gas for both tangled and radial initial magnetic field geometries. Cold gas formation is hardly suppressed with anisotropic thermal conduction because all scales perpendicular to the magnetic field direction are thermally unstable, and long overdense filaments aligned along the local magnetic field condense out of the hot phase (see Fig. \ref{fig:2D}).

We have verified that the simulations with anisotropic thermal conduction show magnetic fields preferentially aligned perpendicular to the radial direction outside the core because of the HBI. The core magnetic field is tangled because of motions stirred by local thermal instability. Simulations without conduction or with isotropic conduction do not show such a bias. This can be seen at large radii in Figure \ref{fig:2D}.

 \begin{figure}
\centering
\psfrag{B}[cc][ ][1][0]{$\dot{M}_{\rm cold} (M_\sun {\rm yr}^{-1})$}
\psfrag{A}{$K_0 ({\rm keV cm}^2$)}
\includegraphics[scale=0.48]{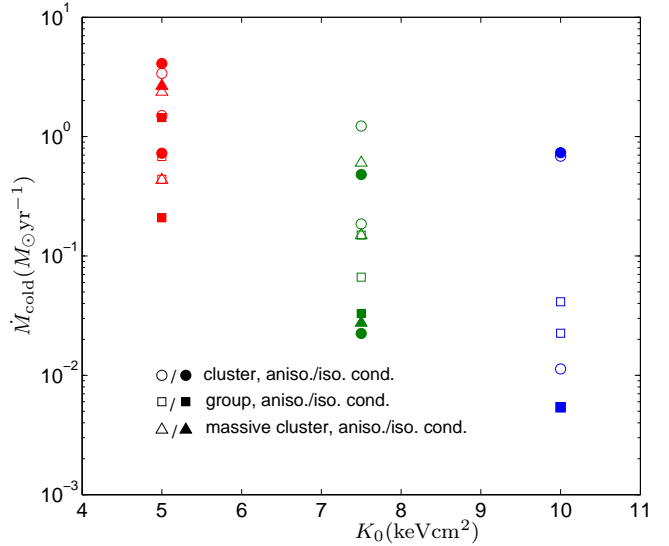}
\caption{The average mass accretion rate in the cold phase ($<0.01$ keV) as a function of core entropy in 2-D runs ($512 \times 256$) with isotropic and anisotropic thermal conduction. The squares, circles and triangles represent group (halo mass $3.8\times 10^{13} \msun$), cluster (halo mass $3.8 \times 10^{14} \msun$) and massive cluster (halo mass $10^{15} \msun$) runs, respectively. Different colors correspond to different values of core entropy. The filled symbols are for isotropic conduction runs with conductivity factors of $f=0.3,~1$ (mass accretion rate decreases with increasing Spitzer fraction). The open symbols are for anisotropic conduction runs with conductivity factors of $f=1,~3$.}
\label{fig:isocond}
\end{figure}
 
Figure \ref{fig:isocond} shows the mass accretion rate in the cold phase for different mass halos and using different Spitzer fractions, with both isotropic (filled symbols) and anisotropic (open symbols) thermal conduction. The mass accretion rate decreases with increasing core entropy and no cold gas is formed for $K_0 > 10$ keV cm$^2$. More importantly, no cold gas is produced in group and massive cluster runs for $K_0 \geq 7.5$ keV cm$^2$  runs using isotropic conduction at the Spitzer value. Observationally, there is not much dependence of the presence of cold gas filaments  (e.g., traced by H$\alpha$) with the halo mass (\citealt{mcd11}; also Fig. 9 in \citetalias{sha12a}). Improving the accuracy of this observational correlation will help constrain conduction in cluster cores. The mass accretion rate in the cold phase is quite insensitive to the halo mass and the Spitzer fraction for runs using anisotropic conduction; the spread in open symbols is much smaller than in filled symbols in Figure \ref{fig:isocond}.
 
 \section{Astrophysical Implications}
 
 This paper takes the important step of including magnetic fields and thermal conduction in studying the condensation of cold filaments in globally stable but locally thermally unstable cores of galaxy clusters. The formalism developed by \citet{mcc12} and \citetalias{sha12a} is based on the observed rough  thermal balance in cluster cool cores, and is able to quantitatively account for the presence of extended cold filaments observed in most cool clusters.  We show that anisotropic thermal conduction at the Spitzer value does not significantly affect the formation of cold gas in our models with thermal balance, but models using isotropic conduction show strongly suppressed cold gas condensation (Fig. \ref{fig:cold_fr}). The simulations with anisotropic conduction show cold gas whenever $t_{\rm cool}/t_{\rm ff} \lesssim 10$ (corresponding to $K_0 \lesssim$ 10 keV cm$^2$) but formation of substantial cold gas with isotropic thermal conduction requires $t_{\rm cool}/t_{\rm ff} < 6$ ($K_0 < 5$ keV cm$^2$). Thus, observations suggest that either thermal conduction is anisotropic or, if conduction is isotropic, it is suppressed by a factor $> 3$ compared to the Spitzer value.
   
 Energy transport to core due to anisotropic thermal conduction is $\lesssim 10\%$ of radiative losses (corresponding value is $\sim 50\%$ for isotropic conduction runs) for cool cores with $K_0 \lesssim 10$ keV cm$^2$. This is due to HBI which aligns magnetic fields perpendicular to the radial direction in presence of anisotropic conduction. Our HBI-suppressed value using realistic field strengths and geometries is consistent with \citealt{ava13}, but clearly shows that AGN feedback is necessary to prevent catastrophic cooling in cores. 
  
 Voit and collaborators (\citealt{voi08,voi11}) have argued, by comparing core radius and the Field length, that cool core clusters show cold multiphase filaments because Field length is smaller than the core radius. While this idea is plausible, it ignores the interplay of local thermal instability and gravity, which suppresses the growth of thermal instability. This interplay is encapsulated in the $t_{\rm cool}/t_{\rm ff} \lesssim 10$ criterion of \citetalias{sha12a}. \citetalias{sha12a} shows that no cold gas condenses from the hot phase even without conduction (so Field length is zero) if $t_{\rm cool}/t_{\rm ff} \gtrsim 10$ in global thermal balance; i.e., the Field length criterion of Voit et al. is not a sufficient condition. Field criterion is also not a necessary condition because all scales perpendicular to the local magnetic field direction are locally thermally unstable with anisotropic conduction, and this is evidenced by the lack of influence of anisotropic thermal conduction on the condensation of cold gas from the hot ICM (Figs. \ref{fig:cold_fr} \& \ref{fig:isocond}). 
 
 It is a coincidence that both $t_{\rm cool}/t_{\rm ff}$ and the Field criteria give similar core entropy ($K_0$) for cluster mass halos. Comparing multiphase cooling in different halos should help distinguish between these two criteria. One can also test this physics in other systems where cooling, heating and gravity are key players such as the base of the solar corona. \citet{sha13} has shown that the ratio $t_{\rm cool}/t_{\rm ff}$ can become $\lesssim 10$ at radii $ \lesssim 1.01 R_\odot$ and this ($\sim 0.01 R_\odot$) is the scale at which phenomena related to cooling, such as coronal rain are observed (note that strong coronal magnetic fields play a major role in shaping these structures; e.g., \citealt{ant12}). In contrast, the Field length for the coronal plasma is $\sim 2 R_\odot$, much bigger than the scales of observed condensations. 
 
The interplay of anisotropic thermal conduction and turbulence has been implicated for the observed bimodality (in terms of core entropy and cooling times; \citealt{cav09,pra10}) of galaxy clusters (\citealt{par10,rus10}). The idea is that the HBI shuts off thermal conduction when turbulence is weak and is unable to overcome effective stable stratification due to HBI, leading to a global catastrophe checked by strong AGN feedback. On the other hand if turbulence is strong ($\gtrsim 100$ km s$^{-1}$), magnetic field lines are isotropic and allow significant heat flux into the core leading to thermal balance without significant AGN feedback. While this is a promising explanation,  it is not the only one. One can imagine the observed bimodality as a natural outcome of cold mode feedback (\citealt{piz05,gil12}). Initially the cooling time is short and it triggers massive cold accretion. Excessive feedback heating raises entropy and the cooling time of the ICM, and quenches massive accretion in the cold phase.  Slowly, after a long cooling time, the cooling cycle can begin again. Thus in this model, clusters are expected to show a bimodal core entropy distribution, corresponding to a cool core state prone to forming multiphase gas and a non-cool-core state with a long cooling time (see, e.g., Fig. 15 in \citetalias{sha12a}). Thus, the observed bimodality does not necessarily constrain thermal conduction and turbulence in the ICM.
 
The sharpness of the observed cold fronts, buoyant bubbles and small scale Kelvin-Helmholtz rolls observed in galaxy clusters argue for a suppressed/anisotropic  thermal conduction (e.g., \citealt{zuh13,kom13}). In this paper we show that the observations of cold gas in different mass halos can also constrain global thermal conduction in cores of galaxy clusters. A comparison of our simulations with observations of extended cold gas in cluster cores clearly rule out isotopic thermal conduction close to the Spitzer value. Anisotropic thermal conduction with Spitzer conductivity is consistent with observations. Future observations of cold gas in different mass halos will further constrain the ICM  conductivity.

 \section*{Acknowledgments}
The numerical simulations were carried out on the computer cluster provided by the start-up grant of PS at IISc. This work is partly supported by the DST-India grant no. Sr/S2/HEP-048/2012. PS acknowledges travel support from the APS ITGAP award, and the local support provided at UC Berkeley by Prof. Quataert. MM is partly supported by NASA ATP grant NNX10AC95G, Chandra theory grant TM2-13004X, and the Thomas and Alison Schneider Chair in Physics at UC Berkeley.
 
 \label{lastpage}

\end{document}